\documentclass[showpacs, pra,twocolumn,preprintnumbers ,amsmath, amssymb, superscriptaddress, aps]{revtex4-2}
\usepackage[normalem]{ulem}
\usepackage{color}
\usepackage{amsmath,amssymb}
\usepackage{pifont}
\usepackage{bbold}
\usepackage{float}
\usepackage{subfloat}
\usepackage[caption=false]{subfig}
\usepackage{tikz}
\usepackage{makecell}
\usepackage{pifont}
\usepackage{graphicx}
\graphicspath{{Figures/}}
\usepackage{dcolumn}
\usepackage{bm}
\usepackage{multirow}
\usepackage{placeins}
\usepackage[colorlinks]{hyperref}
\usepackage{mathtools}
\usepackage{appendix}
\usepackage{caption}
\usepackage{ragged2e}

\def \be{\begin{align}}
	\def \ee{\end{align}}
\def \bea{\begin{eqnarray}}
	\def \eea{\end{eqnarray}}

\begin{document}
	
	\title{
		Gate-tunable spin-valley transport via carrier velocity in monolayer  WSe$_2$}

\author{Otman Bouladiane}
\affiliation{Laboratory of Theoretical Physics, Faculty of Sciences, Choua\"ib Doukkali University, PO Box 20, 24000 El Jadida, Morocco}

\author{Hocine Bahlouli}
\affiliation{Physics Department and IRC Advanced Materials$,$
King Fahd University
of Petroleum $\&$ Minerals$,$
Dhahran 31261$,$ Saudi Arabia}
\author{Clarence Cortes}
\affiliation{Vicerrector\'ia de Investigaci\'on y Postgrado, Universidad de La Serena, La Serena 1700000, Chile}  
\author{David Laroze}
\affiliation{Instituto de Alta Investigación, Universidad de Tarapacá, Casilla 7D, Arica, Chile}
\author{Ahmed Jellal}
\email{a.jellal@ucd.ac.ma}
\affiliation{Laboratory of Theoretical Physics, Faculty of Sciences, Choua\"ib Doukkali University, PO Box 20, 24000 El Jadida, Morocco}

\begin{abstract}

We theoretically investigate spin- and valley-resolved quantum transport in monolayer tungsten diselenide (WSe$_2$) described by an effective massive Dirac Hamiltonian. Particular attention is devoted to a finite barrier region characterized by simultaneously modulated Fermi velocity and scalar potential. The barrier velocity $v_2$ is related to the external velocity $v_1$ through a velocity ratio $\xi=v_2/v_1$, motivated by an optical analogy with the Snell--Descartes law. {The exact refraction condition depends on the full spin- and valley-resolved dispersion, and the simple ratio $\xi=v_2/v_1$ is recovered only in the massless, symmetric limit}. The interplay of intrinsic spin--orbit coupling in the conduction and valence bands, quantified by $\lambda_c$ and $\lambda_v$, with spin- and valley-dependent Zeeman fields, $M_s$ and $M_v$, gives rise to substantial changes in the quasiparticle dispersion, leading to pronounced modifications of the transport characteristics. By solving the Dirac equation and enforcing {current-conserving matching} conditions at the interfaces, we compute the spin- and valley-dependent transmission probability and conductance. Our results demonstrate that the barrier velocity, scalar potential, incidence angle, incident energy, and barrier width serve as effective control parameters for transport, giving rise to strong anisotropy and resonant tunneling features. Furthermore, we show that both the magnitude and orientation of spin- and valley-polarized currents can be continuously tuned via velocity and potential modulation. These findings establish combined velocity and potential engineering as a powerful theoretical framework for controlling spin--valley physics in two-dimensional transition-metal dichalcogenides.

\end{abstract}
\pacs{72.80.Vp, 73.23.-b, 78.67.-n \\
{\sc Keywords}: Monolayer WSe$_2$, velocities, spin-orbit coupling, Zeeman field, transmission, spin-valley, conductance, polarization.}
\maketitle
\section{Introduction}
\label{Intro}
Understanding the behavior of charge carriers at the nanoscale is a central challenge in contemporary condensed matter physics and nanotechnology. Addressing this challenge requires versatile and powerful theoretical frameworks, among which quantum transport plays a pivotal role \cite{Rita2005,Nicholas1980}. Quantum transport phenomena form a cornerstone of modern physics and underpin a wide range of developments in condensed matter physics, mesoscopic systems, and nanoscale device engineering. In particular, simple model systems, such as transport across a single potential barrier, provide valuable insight into fundamental quantum interference mechanisms, including Fabry–Pérot resonances—while simultaneously offering guiding principles for the design of next-generation electronic devices \cite{Martino,Cheol}. The rapid emergence of two-dimensional (2D) materials has further expanded the scope of quantum transport studies, opening new avenues in both materials science and condensed matter physics due to their exceptional and highly tunable mechanical and electronic properties \cite{2D materials-1,2D materials-2}. The experimental isolation of graphene in 2004 marked a watershed moment in this field  \cite{Novoselov}, revealing a material with extraordinary electronic, mechanical, and thermal characteristics \cite{Novoselov2007, approx, Castro2009, absor}. This discovery has since spurred intensive research efforts aimed at identifying and engineering new 2D materials with diverse physical properties and functionalities \cite{Claire Berger, XU}. Several other types of 2D materials have been discovered, such as silicene \cite{silicene-1}. Recently, other types of 2D  {materials} have attracted considerable interest, known as transition metal {dichalcogenides} (TMDs) \cite{TMDCs-1, TMDCs-2, band gap}. Their chemical formula is MX$_2$ (transition metal M = Mo, W, and chalcogen X = S, Se, Te), like graphene, they have a hexagonal crystalline structure \cite{TMDCs-3, TMDCs-4, TMDCs-1}. Many of them are semiconductors, and they exhibit a wide range of remarkable electronic, optical, and mechanical properties \cite{TMDCs-4, TMDCs-5}. The transition from an indirect to a direct band gap in monolayer TMD semiconductors, such as MoX$_2$ and WX$_2$ (X = S, Se) \cite{band gap}. This band-gap transition underpins the development of next-generation electronic and photonic devices, including high-performance transistors \cite{transistor-1,transistor-2}, photodetectors, and photovoltaics \cite{photovoltaics}. Compared to molybdenum-based TMDs (MX$_2$), WSe$_2$ semiconductors are more abundant, less expensive, and less toxic. In addition, they have unique properties due to their direct band gap \cite{WSe2-1}. These monolayers can be obtained by simple mechanical exfoliation of bulk 2H-WSe$_2$ single crystals \cite{WSe2-2, Zhu2011}. Spin-orbit coupling (SOC) is a strong interaction in WSe$_2$ that, unlike other TMDs \cite{Drew2015, Zhu2011}, with an inversion symmetry breaking, induces significant spin-valley coupling. This  makes it possible to design systems that can be used to control the degrees of freedom of spin and valley. In these materials, fermions can acquire a valley pseudo-spin that is very similar to real spin and is associated with a magnetic moment connected to valley-dependent circular dichroism. As a result, it contributes to a range of applications, including the optical generation of valley polarization and quantum coherence between valleys \cite{valley-1}. An external magnetic field can manipulate the degree of freedom of the pseudo-spin valley, increasing degeneracy.

We study quantum transport of {massive} Dirac fermions in monolayer WSe$_2$ across a single potential barrier. The system includes a spatially varying Fermi velocity. Such velocity modulation strongly affects the propagation of Dirac waves. It also modifies the transmission through the barrier. We analyze spin- and valley-dependent transmission, conductance, and polarization. These quantities provide direct information on the transport properties of the system. They show how the electronic states respond to changes in the Fermi velocity inside the barrier. They also describe how waves propagate across regions with different velocities. Control of the spin ($\uparrow,\downarrow$) and valley ($\bm{K}, \bm{K'}$) degrees of freedom is achieved through strong spin–orbit coupling. This coupling acts in both the conduction and valence bands. It is characterized by the parameters $\lambda_c$ and $\lambda_v$. The energy splitting of the circularly polarized components $\sigma_+$ and $\sigma_-$ lifts the degeneracy between the two valleys. The Zeeman effect is also included in our model. It enhances the separation between spin and valley states. This allows a clear distinction between their individual contributions. As a result, spin- and valley-polarized currents can be generated and modulated. Using an analogy with Snell–Descartes law in optics \cite{Concha2010}, we define the velocity ratio $\xi = v_2 / v_1$. {In this} massive, spin- and valley-dependent Dirac problem, the exact relation between the incident angle $\phi^{\tau s_z}$ and the refraction angle $\theta^{\tau s_z}$ is determined by the full local dispersion. The simple ratio $\xi=v_2/v_1$ therefore provides only an approximate description of wave-vector matching at the interfaces \cite{Concha2010}. This ratio compares the Fermi velocity inside the barrier with that outside. By tuning the velocity $v_2$, the transmission of Dirac fermions can be controlled. This tuning depends on spin and valley indices. Our results show that velocity modulation, combined with spin–orbit coupling and Zeeman effects, provides an efficient way to control electronic transport in monolayer WSe$_2$. This approach offers a promising platform for spintronic and valleytronic devices \cite{TMDCs-2,Drew2015}.

The paper is organized as follows. In Section \ref{Theory}, we present the theoretical model  describing electronic transport in a monolayer WSe$_2$. We derive the corresponding energy spectrum and eigenspinors in each region of the system. Section \ref{Tunneling} is devoted to the calculation of the spin- and valley-dependent transmission, conductance, and polarization by applying appropriate boundary conditions at the interfaces $x=0$ and $x=L$. In Section \ref{Results}, we discuss numerical results for the spin- and valley-resolved transport properties as a function of key parameters, including the Fermi velocity, the barrier height, width,  the incident angle. Finally, Section \ref{conclusion} summarizes the main findings of our work.

\hspace{0.1mm}
\section{Theoretical Model}
\label{Theory}
\begin{figure}[htp!]
\centering
\includegraphics[scale=0.5]{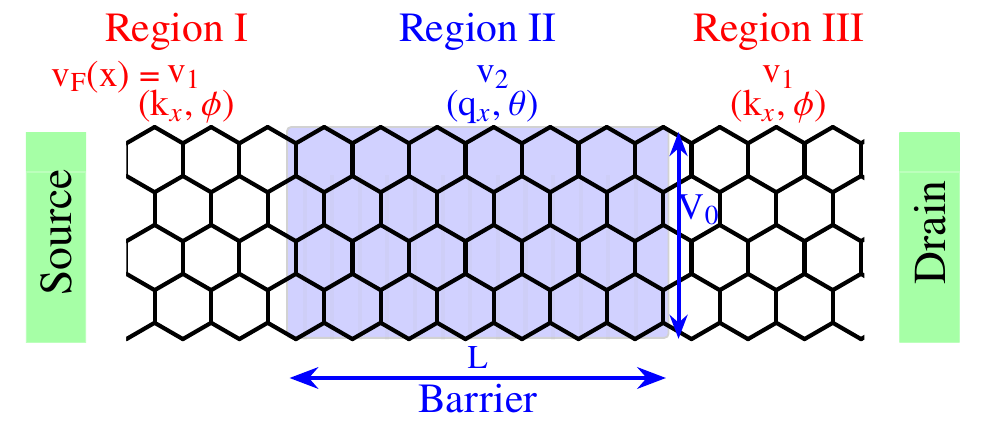}
\caption{\justifying Schematic of the device based on a monolayer  WSe$_2$, where the tunneling setup consists of three regions: region I (source), where the incident carriers are characterized by velocity $v_1$ and angle $\phi$ (regions I and III); region II, corresponding to a potential barrier of height $V_0$ and width $L$ with the velocity $v_2$ and angle $\theta$; and region III (drain), which retains the same parameters as region I.}
\label{schema}
\end{figure}

A single potential barrier is engineered within a monolayer  WSe$_2$. Experimentally, this can be realized by depositing the material onto a suitable substrate and introducing an external electrostatic gate or local doping, as shown schematically in Fig.~\ref{schema}. 
This configuration allows for precise control over both the height and width of the potential barrier, thereby providing an ideal platform for investigating spin–valley-dependent tunneling effects in transition-metal dichalcogenides (TMDs).

Following the low-energy model, we can define the Hamiltonian describing electron motion in WSe$_2$  for both valleys ($\bm{K},\bm{K'}$) and spins ($\uparrow$, $\downarrow$) as \cite{Tahir, Raoux2010}
\begin{widetext}
	\begin{equation}\label{ham}
		H = \frac{1}{2}\{v_F(x),p_x\}\,\tau \sigma_x + v_F(x)\,\sigma_y p_y + \frac{\Delta}{2}\sigma_z + \lambda_c \tau s_z(\sigma_0+\sigma_z) + \lambda_v \tau s_z(\sigma_0-\sigma_z) + (s_zM_s-\tau M_v + V(x))\sigma_0
	\end{equation}
\end{widetext}
where the anticommutator $\{A,B\}=AB+BA$ ensures that the kinetic term containing the position-dependent Fermi velocity is Hermitian \cite{Concha2010} because $v_F(x)$ depends only on $x$, it commutes with $p_y$ and no symmetrization is required there. Equivalently, one may write the kinetic term as $\sqrt{v_F(x)}\,p_x\sqrt{v_F(x)}\,\tau\sigma_x$ \cite{Concha2010}. Introducing the auxiliary spinor $\Phi(x)=\sqrt{v_F(x)}\,\Psi(x)$, the stationary Schr\"odinger equation $H\Psi=E\Psi$ transforms, in each homogeneous region, into the same pair of coupled equations as obtained from the naive Hamiltonian but with $\Psi$ replaced by $\Phi$ and the velocity $v_i$ moved to the effective potential terms. Consequently, the eigenspinors for $\Phi$ retain the functional forms shown below, and the wave vectors $k_x^{\tau s_z}$ and $q_x^{\tau s_z}$ are unchanged. The crucial difference is that $\Phi$, not $\Psi$, must be continuous at the interfaces \cite{Concha2010}.
Here, $p_x$, $p_y$ are the in-plane components of the electron momentum (with $p_{x,y}=-i \hbar \partial_{x,y}$),  $\sigma_i$ ($i = x, y, z$) represent the Pauli matrices operating in pseudo-spin space, and $\sigma_0$ being the $2 \times 2$ identity matrix. The parameter $\tau = \pm 1$ denotes the non-equivalent valleys ($\bm{K},\bm{K'}$), while $s_z = \pm 1$ corresponds to the spin up and down states, respectively.
The term $\Delta =1.7$ eV represents the intrinsic direct band gap of WSe$_2$, and {$\lambda_c \approx 7.5$ meV, $\lambda_v \approx 112.5$ meV account for the spin-orbit coupling strength in the conduction and valence bands, respectively \cite{Tahir, Zhu2011, PhysRevB.107.245407}.
The parameters $M_s$ and $M_v$ represent the effective spin and valley splitting energies, respectively. They are introduced in the form of Zeeman-like terms,
$M_s=g'\mu_B B/2$ and $M_v=g_v\mu_B B/2$, where
$g'=g'_e+g'_s$ is the effective spin $g$-factor. In particular,
$g'_e=2$ is the free-electron $g$-factor, while
$g'_s=0.21$ describes the out-of-plane contribution originating from the strong spin--orbit coupling in $\mathrm{WSe_2}$. The valley splitting is characterized by the valley $g$-factor $g_v=4$ \cite{valley-1,Srivastava2015}.
{We emphasize that the values $M_s=30~\mathrm{meV}$ and $M_v=60~\mathrm{meV}$ (as adopted in \cite{Tahir}) are not intended to represent the bare Zeeman splittings produced by a conventional external magnetic field alone. Indeed, the Zeeman energy is given by $E_Z=g\mu_B B$, which is only of the order of a few meV even for strong laboratory fields. These values are therefore considered as effective spin- and valley-dependent energy scales describing enhanced exchange effects, which may arise, for example, from proximity coupling with ferromagnetic substrates such as EuS \cite{Zhao2019,valley-1}, electric-field-induced band modifications, or strain-engineered valley splitting. Such effective descriptions are widely used in studies of spintronic and valleytronic devices \cite{Zhao2019}. Alternatively, these parameters can be regarded as phenomenological quantities capturing the combined influence of exchange interactions and spin–orbit coupling.}
The modulated Fermi velocity $v_F(x)$ is \cite{Raoux2010}
\begin{align}\label{vF}
v_F(x)=
\begin{cases}
	v_1, & \text{I } x<0\\
	v_2, & \text{II } 0<x<L\\
	v_1, & \text{III } x>L
\end{cases}
\end{align}
with $v_1 = v_F = 5\times10^5$ m/s, and $v_2$ in region II is adjustable.
 {The velocity profile adopted here is motivated by experimentally accessible mechanisms rather than being only a phenomenological assumption. A spatially varying Fermi velocity can be realized in two-dimensional Dirac materials. One route is to modify the dielectric environment of the sample, since dielectric screening affects the electronic interactions and consequently the renormalized Fermi velocity, as measured for graphene on different substrates \cite{Hwang2012}. A second route is to place a metallic layer close to part of the sheet, where its modified screening properties can induce a local renormalization of the Fermi velocity \cite{Raoux2010}. Applying such a modification to the central region alone can be modeled by the step profile $v_1 \to v_2 \to v_1$ considered in this work, with $v_2$ below or above $v_1$ depending on whether the local screening environment is stronger or weaker.}
As {depicted} in Fig.~\ref{schema}, the  single rectangular barrier along $x$-direction is  
\begin{equation}
V(x)=	\begin{cases}
	V_0 & 0<x<L\\
	0 &  \text{Otherwise}
\end{cases}
\end{equation}
where $V_0$ is the  barrier height and $L$ is its width. We consider the width $L_y$ (along the $y$-direction) to be much larger than the {barrier} region length $L$ (along the $x$-direction). Under this condition, edge effects can be neglected, and fermions are assumed to retain translational invariance along the $y$-axis. Consequently, their dynamics are effectively governed by motion along the $x$-direction.  Due to the symmetry of the system, we can split the eigenspinors of Eq. \eqref{ham} into two components
$\Psi=(\psi_c,\psi_v)^\dagger e^{ik_yy}e^{-iEt/\hbar}$.

We solve the time-independent Schrödinger equation $H \Psi = E \Psi$ in order to obtain a general solution. As a result, we obtain 
\begin{widetext}
\begin{align}
	v_F(x)\hbar \left[-i \tau \frac{\partial}{\partial x} - i   k_y  \right]\phi_v(x) &= \left[ E^{\tau s_z} - V(x) -\frac{\Delta}{2}- 2 \lambda_c \tau s_z-s_zM_s+\tau M_v \right] \phi_c(x) \\
	v_F(x) \hbar \left[-i \tau \frac{\partial}{\partial x} + i k_y   \right]\phi_c(x) &= \left[ E^{\tau s_z} - V(x) +\frac{\Delta}{2}-2\lambda_v \tau s_z-s_zM_s+\tau M_v \right] \phi_v(x).
\end{align}
\end{widetext}
As a result, 
{the corresponding eigenspinors in the three regions are obtained} for a given valley $\tau$ and spin $s_z$ {indices} 
\begin{widetext}
\begin{align}
	&\Psi^{\tau s_z}_{\text{I}}(x,y) = 
	\left[
	\begin{pmatrix} 1 \\ \gamma^{\tau s_z} e^{i\phi^{\tau s_z}} \end{pmatrix} e^{i k_x^{\tau s_z}x}
	+ r^{\tau s_z}\begin{pmatrix} 1 \\ - \gamma^{\tau s_z} e^{-i\phi^{\tau s_z}} \end{pmatrix} e^{-i k_x^{\tau s_z}x}
	\right]e^{ik_yy}\\
	&
	\gamma^{\tau s_z} = \left(\frac{ E^{\tau s_z} -\frac{\Delta}{2}- 2 \lambda_c \tau s_z-s_zM_s+\tau M_v }{E^{\tau s_z} + \frac{\Delta}{2}-2\lambda_v \tau s_z-s_zM_s+\tau M_v }\right)^\frac{1}{2}
	\\
	&\Psi^{\tau s_z}_{\text{II}}(x,y) =
	\left[
	a\begin{pmatrix} 1 \\ \delta^{\tau s_z}  e^{i\theta^{\tau s_z}} \end{pmatrix} e^{i q_x^{\tau s_z}x}
	+ b\begin{pmatrix} 1 \\ - \delta^{\tau s_z} e^{-i\theta^{\tau s_z}} \end{pmatrix} e^{-i q_x^{\tau s_z}x}
	\right]e^{ik_yy}\\
	&\delta^{\tau s_z} = \left( \frac{ E^{\tau s_z} -V_0-\frac{\Delta}{2}- 2 \lambda_c \tau s_z-s_zM_s+\tau M_v}{E^{\tau s_z} -V_0 + \frac{\Delta}{2}-2\lambda_v \tau s_z-s_zM_s+\tau M_v }\right)^\frac{1}{2}
	\\
	& \Psi^{\tau s_z}_{\text{III}} (x,y)=
	t^{\tau s_z}\begin{pmatrix} 1 \\ \gamma^{\tau s_z} e^{i\phi^{\tau s_z}} \end{pmatrix}
	e^{i k_x^{\tau s_z}x}e^{ik_yy}
\end{align}
\end{widetext}
where $r^{\tau s_z}$ and $t^{\tau s_z}$ are the spin- and valley-dependent reflection and transmission coefficients, $a$ and $b$ are two constants. 
The wave vectors $k_x^{\tau s_z}$ and $q_x^{\tau s_z}$ along $x$ can be expressed in terms of energy $E^{\tau s_z}$ as
\begin{widetext}
\begin{align}
	&k_x^{\tau s_z} =\tau \sqrt{\frac{1}{\hbar^2 v_1^2}( E^{\tau s_z}-\frac{\Delta}{2}- 2 \lambda_c \tau s_z-s_zM_s+\tau M_v)(E^{\tau s_z} +\frac{\Delta}{2}-2\lambda_v \tau s_z-s_zM_s+\tau M_v)-k_y^2} \label{kx}\\
	&q_x^{\tau s_z} =\tau \sqrt{\frac{1}{\hbar^2 v_2^2}( E^{\tau s_z} - V_0 -\frac{\Delta}{2}- 2 \lambda_c \tau s_z-s_zM_s+\tau M_v)(E^{\tau s_z} - V_0+\frac{\Delta}{2}-2\lambda_v \tau s_z-s_zM_s+\tau M_v)-k_y^2} \label{qx}.
\end{align}
\end{widetext}
The spin- and valley-dependent incident and refraction angles are directly determined by particle propagation in each region. These angles are related to the corresponding wave vectors in the different regions of the system by
\begin{align}
\phi^{\tau s_z}=\tan^{-1}\left(\frac{k_y}{\tau k_x^{\tau s_z}}\right), \quad	\theta^{\tau s_z}=\tan^{-1}\left(\frac{k_y}{\tau q_x^{\tau s_z}}\right).
\end{align}
It is important to emphasize that in this massive, spin- and valley-dependent Dirac problem, the refraction condition is not simply $\sin\theta/\sin\phi=v_2/v_1$. From the conservation of transverse momentum $k_y$ and the local dispersion in each region, the exact generalized Snell law reads \cite{Concha2010}
\begin{equation}
	\frac{\sin\theta^{\tau s_z}}{\sin\phi^{\tau s_z}} = \xi
	\sqrt{\frac{\left(E-M_{cc}^{\rm I}\right)\left(E-M_{vv}^{\rm I}\right)}{\left(E-V_0-M_{cc}^{\rm I}\right)\left(E-V_0-M_{vv}^{\rm I}\right)}}
\end{equation}
where $M_{cc}^{\rm I}=\frac{\Delta}{2}+2\lambda_c\tau s_z+s_zM_s-\tau M_v$ and $M_{vv}^{\rm I}=-\frac{\Delta}{2}+2\lambda_v\tau s_z+s_zM_s-\tau M_v$ are the effective mass terms in region I. The simple ratio $\xi=v_2/v_1$ is recovered only when the band gap, spin--orbit couplings, and potential offset vanish \cite{Concha2010}.

The spin- and valley-dependent energy spectrum is obtained by diagonalizing the Hamiltonian given in Eq.~\eqref{ham}. The resulting eigenvalues depend explicitly on the wave vector $k$ and on the position-dependent Fermi velocity $v_F(x)$. This dispersion relation reflects the combined effects of the band gap, spin–orbit coupling, and spin- and valley-dependent exchange interactions. As a consequence, the energy spectrum exhibits distinct branches associated with different spin and valley indices, which play a central role in determining the transport properties of the system. This is
\begin{widetext}
\begin{align}\label{Ener}
	&E^{\tau s_z}_s = V(x) + s_zM_s-\tau M_v + \tau s_z (\lambda_c + \lambda_v) + s \hbar v_F(x)\sqrt{k^2 + \left[\frac{\Delta + 2 \tau s_z (\lambda_c - \lambda_v)}{2 \hbar v_F(x)}\right]^2}.
\end{align}
\end{widetext}
The sign $s = \pm$ corresponds to the conduction ($+$) and valence ($-$) bands. The wave vector is given by $k^{\tau s_z} = \sqrt{(k_x^{\tau s_z})^2 + k_y^2}$ in regions (I, III) and by $k^{\tau s_z} = \sqrt{(q_x^{\tau s_z})^2 + k_y^2}$ inside the barrier. This energy spectrum is spin- and valley-dependent, meaning that electrons with different spins or valleys have different energy levels.

\section{Tunneling Characteristics}
\label{Tunneling}
{\bf Transmission:} Because the Hamiltonian is Hermitian, the correct matching conditions follow from the continuity of the auxiliary spinor $\Phi=\sqrt{v_F}\Psi$ at the interfaces. For the physical spinor $\Psi$ this translates into
\begin{align}
	\sqrt{v_1}\,\Psi^{\tau s_z}_{\text{I}}(0) &= \sqrt{v_2}\,\Psi^{\tau s_z}_{\text{II}}(0)\\
	\sqrt{v_2}\,\Psi^{\tau s_z}_{\text{II}}(L) &= \sqrt{v_1}\,\Psi^{\tau s_z}_{\text{III}}(L).
\end{align}
Defining the rescaled barrier coefficients $a'= \sqrt{\xi }a$ and $b'=  \sqrt{\xi }b$, with $\xi =\frac{v_2}{v_1}$, the velocity-dependent prefactors cancel identically at both interfaces. The continuity conditions for $\Phi$ then yield the linear system
\begin{widetext}
	\begin{align}
		&1+r^{\tau s_z}=a'+b'\\
		&\gamma^{\tau s_z} e^{i\phi^{\tau s_z}}- \gamma^{\tau s_z} r^{\tau s_z}e^{-i\phi^{\tau s_z}}=a' \delta^{\tau s_z}  e^{i\theta^{\tau s_z}}-b' \delta^{\tau s_z}  e^{-i\theta^{\tau s_z}}\\
		&t^{\tau s_z}e^{i k_x^{\tau s_z} L}=a'e^{i q_x^{\tau s_z}L}+b'e^{-i q_x^{\tau s_z} L}\\
		&\gamma^{\tau s_z} t^{\tau s_z} e^{i\phi^{\tau s_z}} e^{i k_x^{\tau s_z} L}=a' \delta^{\tau s_z}  e^{i\theta^{\tau s_z}} e^{i q_x^{\tau s_z} L }-b' \delta^{\tau s_z}  e^{-i\theta^{\tau s_z}} e^{-i q_x^{\tau s_z} L }.
	\end{align}
\end{widetext}
This system is formally identical to that obtained from naive spinor continuity, but it is now rigorously derived from the Hermitian Hamiltonian. Since the closed-form solution for $t^{\tau s_z}$ depends only on the algebraic structure of the linear system, we obtain
\begin{widetext}
\begin{align}
	t^{\tau s_z}=
	\frac{e^{-i \tau k_x^{\tau s_z} L }\cos\phi^{\tau s_z}\cos\theta^{\tau s_z}}
	{\cos\phi^{\tau s_z}\cos\theta^{\tau s_z}\cos( q_x^{\tau s_z} L)
		- i \sin( q_x^{\tau s_z} L )(\frac{\gamma^{ \tau s_z 2} + \delta^{\tau s_z 2}}{2 \gamma^{\tau s_z} \delta^{\tau s_z}}- \sin\phi^{\tau s_z}\sin\theta^{\tau s_z})}.
\end{align}
\end{widetext} 
Using the conserved current $j_x=\Psi^\dagger v_F(x)\tau\sigma_x\Psi=\Phi^\dagger\tau\sigma_x\Phi$, one finds 
the transmission $T^{\tau s_z}=|t^{\tau s_z}|^2$ and reflection $R^{\tau s_z}=|r^{\tau s_z}|^2$ probabilities. Explicitly, we have 
\begin{widetext}
\begin{equation}\label{trans}
	T^{\tau s_z} =
	\frac{\cos^2\phi^{\tau s_z} \cos^2\theta^{\tau s_z}}
	{\cos^2\phi^{\tau s_z} \cos^2\theta^{\tau s_z} \cos^2( q_x^{\tau s_z} L)
		+ \sin^2( q_x^{\tau s_z} L)\left(\frac{\gamma^{ \tau s_z 2} + \delta^{\tau s_z 2}}{2 \gamma^{\tau s_z} \delta^{\tau s_z}}-  \sin\phi^{\tau s_z} \sin\theta^{\tau s_z}\right)^2}
\end{equation}
\end{widetext}
where the conservation is satisfied
	\begin{equation}
		R^{\tau s_z}+T^{\tau s_z}=1
	\end{equation}
	for each spin--valley channel, as required by unitarity.
The transmission probability $T^{\tau s_z}$ depends explicitly on the incident and refraction angles, reflecting the phase-coherent nature of resonant tunneling through the barrier structure. It is also spin- and valley-dependent, which will enable selective control of fermionic transport in WSe$_2$. Note that $T^{\tau s_z}(\phi^{\tau s_z}) = T^{\tau s_z}(-\phi^{\tau s_z})$ reflects the mirror symmetry of the system. For normal incidence $\phi^{\tau s_z}\to 0$, the transverse momentum vanishes and Eq.~\eqref{trans} reduces to
\begin{align}
T^{\tau s_z}(0) = \frac{1}{\cos^2( q_x^{\tau s_z} L)
	+ \sin^2( q_x^{\tau s_z} L)\left(\frac{\gamma^{ \tau s_z 2} + \delta^{\tau s_z 2}}{2 \gamma^{\tau s_z} \delta^{\tau s_z}}\right)^2}.
	\end{align}
	Perfect transparency ($T^{\tau s_z}=1$) at normal incidence requires the additional condition $\gamma^{\tau s_z}=\delta^{\tau s_z}$, which is generally not satisfied in the presence of a finite band gap and spin--orbit coupling. In the generic massive case, $T^{\tau s_z}(0)<1$, except at the resonant points $q_x^{\tau s_z}L=n\pi$ where $\sin(q_x^{\tau s_z}L)=0$. When $\gamma^{\tau s_z} =\delta^{\tau s_z} = 1$, which corresponds to the massless limit, the barrier becomes fully transparent for all $L$, reproducing the familiar Klein tunneling phenomenon observed in graphene \cite{Castro2009,klientun}.
	The transmission Eq.~\eqref{trans} clearly shows that the number of oscillations decreases as the Fermi velocity $v_2$ {increases} because fermions propagate with a smaller wave vector $q_x^{\tau s_z}$, which excludes resonance conditions. For the values of $q_x^{\tau s_z} L$, satisfying the relation $q_x^{\tau s_z} L = n \pi$, with $n$ being an integer, we obtain $T^{\tau s_z} = 1$, regardless of the value of $\phi^{\tau s_z}$, therefore the barrier becomes completely transparent.  \\
	
	{\bf Conductance:} 
	Experimental studies typically focus on measuring conductance \(G\). Conductance is a global quantity that relates the total current to the applied voltage drop. In contrast, conductivity $\sigma$ connects the local current density to the electric field. In two-dimensional systems, these two quantities have the same physical units \cite{condu0}.
	To quantify the ease with which electrons propagate through the system under external perturbations, such as a potential barrier, we introduce the concept of quantum conductance. Within this framework, and using the calculated transmission probabilities, the zero-temperature conductance for each spin--valley channel is defined as \cite{butker,conduct1,ELAITOUNI2024}
\begin{align}
	G_{\tau {s_z}} = \frac{G_0}{2\pi} \int_{-\phi^{\text{max}}}^{\phi^{\text{max}}} T^{\tau {s_z}}(E, \phi)\,\cos\phi \, d\phi,
\end{align}
where ${s_z}=\uparrow,\downarrow$ and $\tau=K,K'$. The valley-resolved, spin-resolved, and total conductances are, respectively,
\begin{align}
	G_\tau = \sum_{{s_z}} G_{\tau \sout{s}{s_z}}, \quad G_{{s_z}} = \sum_{\tau} G_{\tau {s_z}}, \quad G_T = \sum_{\tau,{s_z}} G_{\tau {s_z}}.
\end{align}
where  $G_0 = (e^2 L_y)/h$ is the conductance unit \cite{Feng2012}.\\

{\bf Polarization:} 
Spin polarization $P_s$ and valley polarization $P_v$ are fundamental quantities that characterize conductance asymmetries associated with quantum degrees of freedom. Specifically, they quantify the imbalance between contributions from opposite spin states and from different electronic valleys. In terms of the total conductance $G_T$, they are consistently defined as \cite{Feng2012,Rekha2025,Moldovan2012}
\begin{align}
	P_s &= \frac{G_{\uparrow}-G_{\downarrow}}{G_T}=\frac{\sum_{\tau}(G_{\tau\uparrow}-G_{\tau\downarrow})}{G_T},\\
	P_v &= \frac{G_{K}-G_{K'}}{G_T}=\frac{\sum_{s}(G_{K {s_z}}-G_{K' { s_z}})}{G_T}.
\end{align}

Next, we will perform a numerical analysis of the spin- and valley-resolved transmission probability, conductance, and polarization in monolayer WSe$_2$. The calculations are carried out by varying key system parameters, including the Fermi velocity inside the barrier, the barrier height and width, and the angle of incidence. The numerical results reveal pronounced spin- and valley-dependent features in the transmission spectra, which arise from the combined effects of spin–orbit coupling, Zeeman splitting, and velocity modulation. These effects lead to strong modulation of the conductance and enable highly tunable spin and valley polarization. Our analysis demonstrates that appropriate tuning of the barrier and velocity parameters provides an efficient mechanism for controlling spin- and valley-polarized transport in WSe$_2$ monolayers.

\section{Results and discussion}
\label{Results}

\begin{figure}[htp!]
\centering
\includegraphics[scale=0.45]{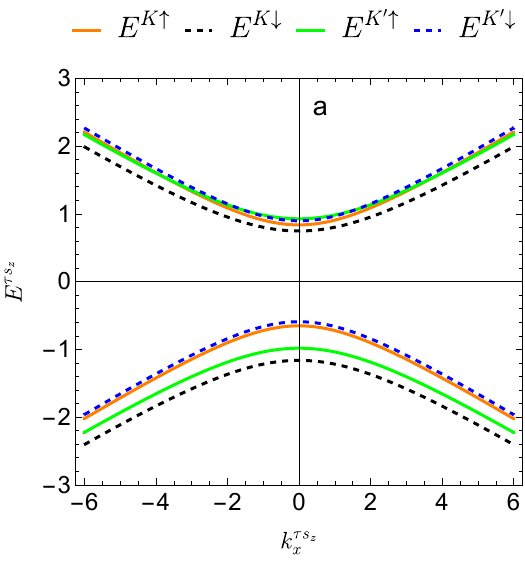}
\includegraphics[scale=0.45]{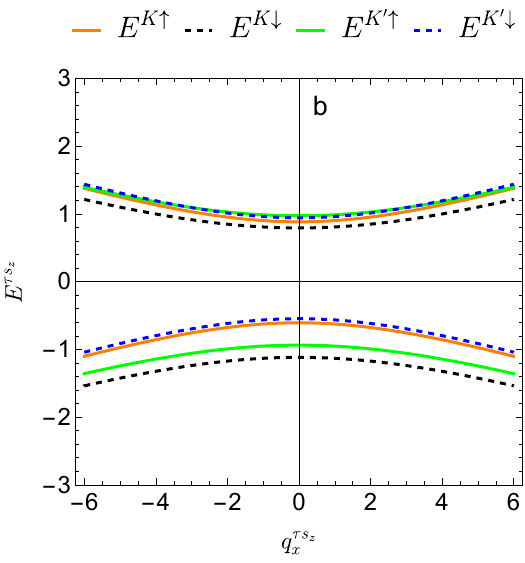}\\ 
\includegraphics[scale=0.45]{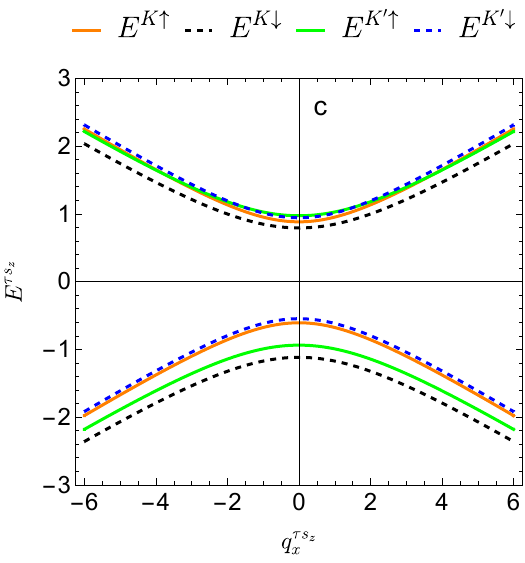}
\includegraphics[scale=0.45]{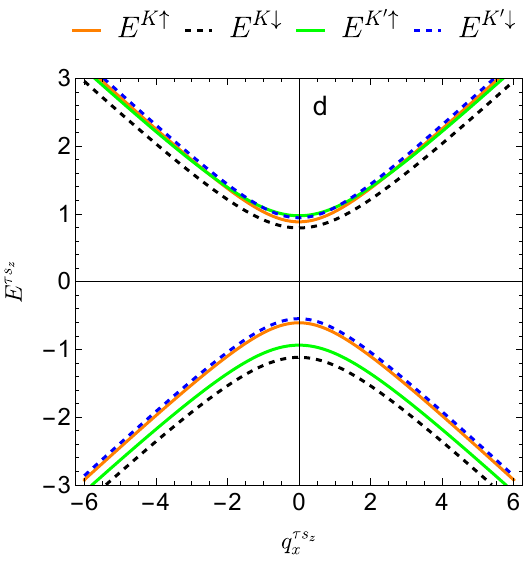}
\caption{\justifying Spin- and valley-dependent energy dispersion $E_s^{\tau s_z}$ as a function of wave vector components $k_x^{\tau s_z}$ and $q_x^{\tau s_z}$ along $x$-direction for $V_0 = 45$ meV,  and $L= 7$ nm. (a): $k_x^{\tau s_z}$  in regions I and  III. (b–d): $q_x^{\tau s_z}$ in region II showing the effect of the velocity ratio (b): $\xi = 0.5$, (c): $\xi = 1$,  and (d): $\xi = 1.5$ {(as adopted in \cite{X.-J.Hao})}  on each energy channel $E_s^{\tau s_z}$.}

\label{Energy}
\end{figure}

Figure~\ref{Energy} shows how energy disperses based on different spin and valley states when the wave-vector reaches various magnitudes. Fig.~\ref{Energy}a displays the electronic dispersion expressed in Eq.~\eqref{Ener} for zero external potential $V(x)=0$, which measures the longitudinal wave-vector component $k_x$. The resulting spectrum consists of eight parabolic bands, with the largest band gap occurring for the \(\bm{K}{\uparrow}\) and \(\bm{K'}{\downarrow}\) states. The inclusion of a Zeeman field breaks the valley symmetry and introduces an energy asymmetry between the \(\bm{K}\) and \(\bm{K'}\) valleys, while the overall electron--hole symmetry of the spectrum is preserved.
Fig.~\ref{Energy}b-d shows the energy spectrum in region~II in the presence of a finite potential \(V_0\) for different values of the Fermi velocity within the barrier. This clearly demonstrates the impact of velocity modulation within the barrier on the transport of massive Dirac fermions in the WSe$_2$ monolayer quantum structure. 
The Fermi velocity only slightly affects the dispersion pattern at high incidence angles. Fig. \ref{Energy}b shows that reducing the Fermi velocity within the barrier makes the bands flatter. As a result, the variation of energy $E_s^{\tau s_z}$ with $q_x^{\tau s_z}$ is weak, which corresponds to a smaller carrier mobility inside the barrier. In contrast, the examination of the outcome shown in Fig.~\ref{Energy}c  indicates that the dispersion becomes more pronounced and approaches that of panel a, with the distinct gap induced by $V_0$. This indicates transport properties closer to those of regions I and III. Finally, in Fig.~\ref{Energy}d, increasing the carrier velocity enhances the band curvature and strengthens the dispersion. In this case, the $\bm{K}\uparrow$ and $\bm{K'}\downarrow$ channels exhibit larger gaps than the remaining channels. 
{An increase in the barrier Fermi velocity $v_2$ results} in high carrier mobility, which causes the energy bands to bend more sharply. The spin--orbit coupling terms further induce spin- and valley-dependent energy shifts, governed by the spin index \(s_z\) and the valley index \(\tau\).

\begin{figure}[htp!]
\centering
\includegraphics[scale=0.418]{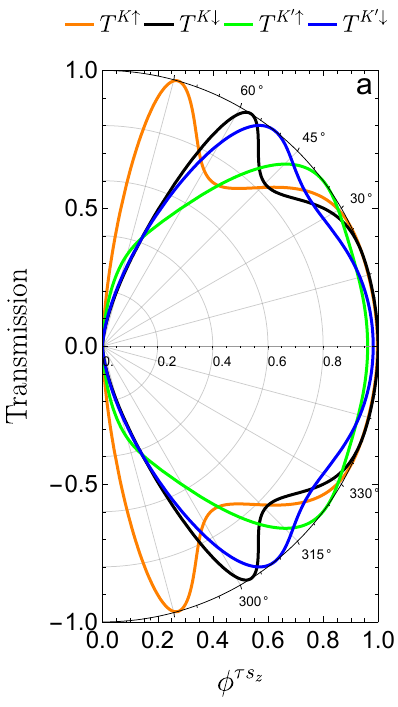}
\includegraphics[scale=0.418]{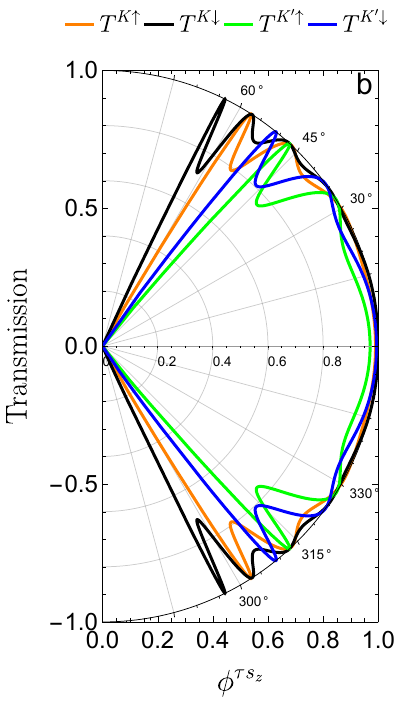}
\includegraphics[scale=0.418]{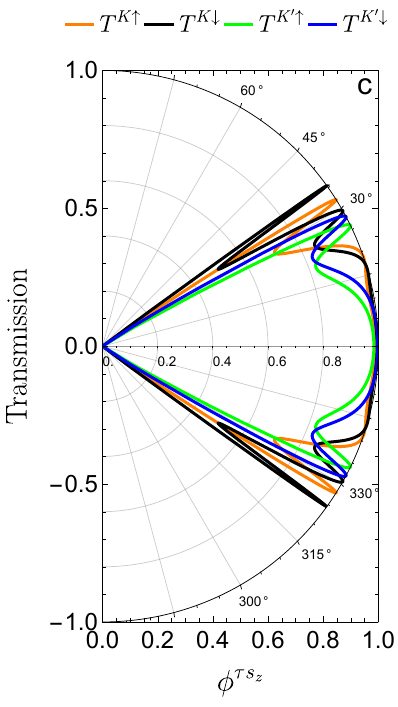}
\caption{\justifying Spin- and valley-dependent transmission  as a function of incident angle $\phi^{\tau s_z}$ for $V_0 = 45$ meV,  $L = 7$ nm,  $E^{\tau s_z} = 1.05$ eV, and three velocity ratio values (a): $\xi = 0.5$, (b): $\xi = 1$,  (c): $\xi = 1.5$. }
\label{T-phi}
\end{figure}

In Fig.~\ref{T-phi}, we present  the transmission $T^{\tau s_z}$ versus the incident angle $\phi^{\tau s_z}$, to demonstrate how the barrier selectively permits specific propagation directions depending on the Fermi velocity within the barrier $\xi$. Fig.~\ref{T-phi}a around normal incidence angles ($\phi^{\tau s_z} \to 0$) the curves imply high transmission.  Pronounced peaks appear around 45$^\circ$ and 60$^\circ$ for $\xi$ = 0.5. Clearly, for these directions, the transmission is almost perfect over a broad band, with visible channel separation. In contrast, Fig.~\ref{T-phi}b, corresponding to $\xi = 1$, exhibits 
{fewer but more pronounced} oscillations, with sharper maxima near 45$^\circ$, while the transmission is reduced in this range. For $\xi = 1.5$  in Fig.~\ref{T-phi}c, the transmission curves compress toward smaller incidence angles. Well-defined peaks appear. Distinct transmission channels display a clear separation between spin and valley contributions, most prominently around an incidence angle of approximately 30$^\circ$, where the differences reach their maximum. 
The mechanism behind the outcome shown in Fig.~\ref{T-phi} is the presence of strong spin-orbit coupling in the conduction and valence bands. This leads to spin-valley dependent energy splitting, as shown in Fig.~\ref{Energy}. Additionally, the Zeeman fields $M_v$ and $M_s$ lift the degeneracy between $\bm{K}$ and $\bm{K'}$, respectively. These interaction terms render certain transport channels propagating while rendering others evanescent. 
 {A decrease} in the Fermi velocity within the barrier 
  {increases} the longitudinal wave vector, $q_x^{\tau s_z}$, pushing the oscillations closer together. This results in finer Fabry-Pérot interference periodicity and denser peaks due to the rapid satisfaction of resonance conditions. In our system, the spin and valley Zeeman field terms result in a weak wave vector for the $\bm{K'}\uparrow$ channel, explaining why it is immediately evanescent.

\begin{figure}[htp!]
\centering
\includegraphics[scale=0.46]{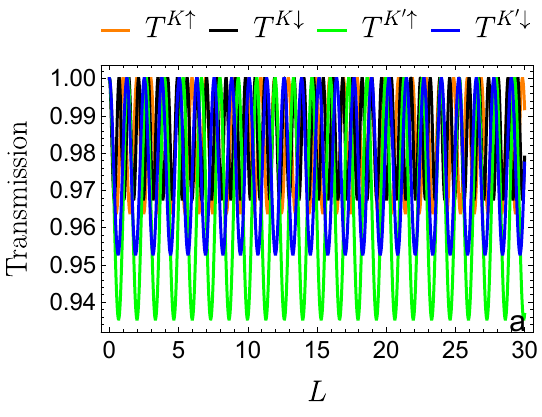}
\includegraphics[scale=0.46]{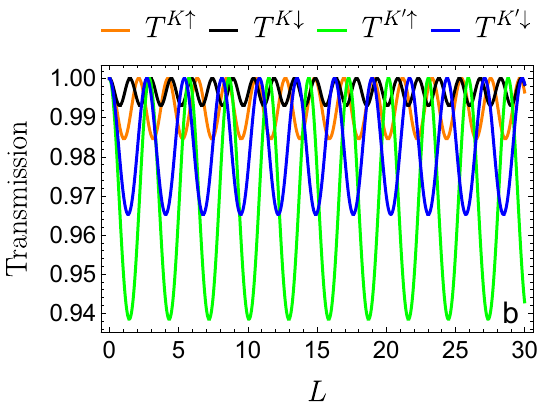}
\includegraphics[scale=0.46]{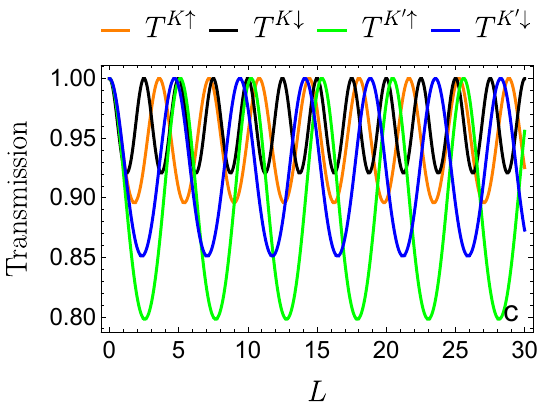}
\caption{\justifying Spin- and valley-dependent transmission  as a function of the barrier width  $L$ for 
$V_0 = 45$~meV, $E^{\tau s_z} = 1.05$ eV, $\phi^{\tau s_z} = 20^\circ$, and
three velocity  ratio values (a): $\xi = 0.5$, (b): $\xi = 1$, (c): $\xi = 1.5$. }
\label{T-L}
\end{figure}

Figure \ref{T-L} shows the  transmission $T^{\tau s_z}$ versus the barrier width $L$ for $V_0 = 45$ meV, $E^{\tau s_z} = 1.05$ eV, and three velocity ratio values $\xi=(0.5,1,1.5)$.  The transmission shows regular periodic oscillations, which are characteristic of Fabry-P\'{e}rot quantum cavities in monolayer WSe$_2$. The large number of oscillations with a small period seen in Fig. \ref{T-L}a can be explained by the small corresponding ratio $\xi=0.5$ associated with a small barrier velocity $v_2$, which gives rise to a larger wave vector inside the barrier $q_x^{\tau s_z}$. Therefore, to achieve the resonance conditions $q_x^{\tau s_z} L = n \pi$, smaller $L$ values are needed. For $\xi =1$,
Fig.~\ref{T-L}b indicates strong adaptation between regions, with less reflection at boundary interfaces. 
Increasing the barrier velocity to $\xi =1.5$ in Fig.~\ref{T-L}c reduces the wave vector $q_x^{\tau s_z}$. This
{reduces the number of oscillations} and increases their period because the resonance conditions are barely satisfied, resulting in more visible spin-valley splitting.
The carriers are able to move more quickly across the potential region because they experience less effective confinement. 
The $\bm{K}\downarrow$ channels show 
{relatively high} transmission amplitudes, 
 {with minima of} $T^{\bm{K}\downarrow} \approx 0.96$ at $\xi = 0.5$ and $T^{\bm{K}\downarrow} \approx 0.9$ at $\xi = 1.5$ according to the results in Fig.~\ref{T-L}a and Fig.~\ref{T-L}c.
  {In contrast, the $\bm{K'}\uparrow$ channel exhibits the deepest minima, reaching $T^{\bm{K'}\uparrow} \approx 0.80$ when the velocity ratio reaches $\xi = 1.5$.} The combination of spin–orbit coupling components ($\lambda_c,\lambda_v$) with Zeeman fields ($M_s, M_v$) creates spin–valley-specific band edge shifts. The shifts create direct impacts on longitudinal wave vectors $k_x^{\tau s_z}$ and $q_x^{\tau s_z}$. Valley channels with larger wave vectors lead to smaller incident angles $\phi^{\tau s_z}$ at a given incident energy $E^{\tau s_z}$. The process leads to an increased phase accumulation ({large} number of oscillations) within the barrier because of the relation $q_x^{\tau s_z} L=n\pi$, which results in stronger constructive interference and higher transmission rates.

\begin{figure}[htp!]
\centering
 \includegraphics[scale=0.46]{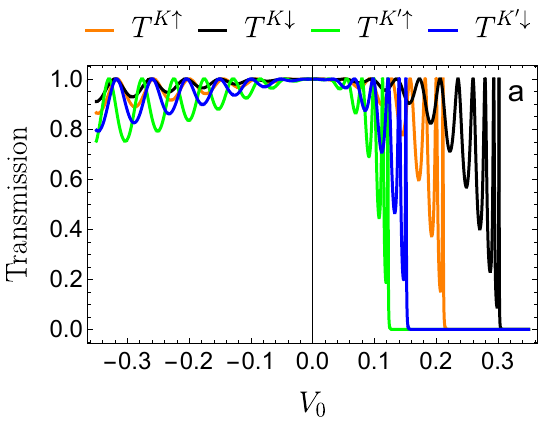}
\includegraphics[scale=0.46]{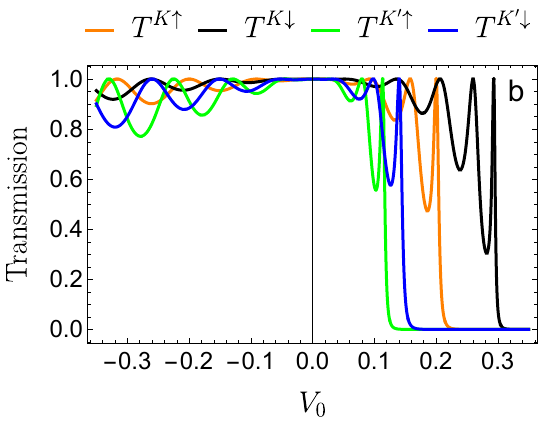}
\includegraphics[scale=0.46]{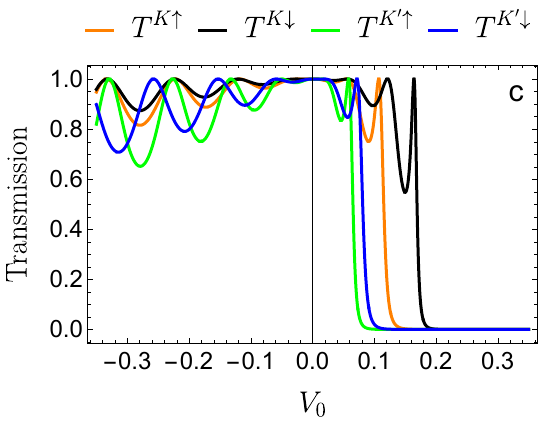}
\caption{\justifying Spin- and valley-dependent transmission  as a function of the barrier height $V_0$ for  $L = 7$ nm,  $E^{\tau s_z} = 1.05$ eV, $\phi^{\tau s_z} = 0^\circ$, and three velocity  ratio values (a): $\xi = 0.5$, (b): $\xi = 1$,  (c): $\xi = 1.5$. }
\label{T-V0}
\end{figure}

Figure~\ref{T-V0} presents the transmission  $T^{\tau s_z}$ versus  the barrier height $V_0$ \sout{for} for  $L = 7$ nm,  $E^{\tau s_z} = 1.05$ eV, and $\xi=(0.5, 1, 1.5)$.  We observe that the oscillation positions and amplitudes depend significantly on the ratio $\xi$. 
Our system exhibits resonant behavior through transmission peaks that reach unity ($T^{\tau s_z} = 1$) at specific potential values.
As seen in Fig.~\ref{T-V0}a with $\xi=0.5$,  a 
{larger} wave vector pushes the oscillations to become 
 {denser} and narrower at lower Fermi velocities within the barrier. For  Fig.~\ref{T-V0}b with $\xi=1$,  we observe more regular, but less frequent, oscillations with a larger period. For the case of  $\xi=1.5$, Fig.~\ref{T-V0}c shows that the effect is more pronounced, with 
 {fewer} oscillations and a visible increase in their period. This makes the effect of velocity $v_2$ on transmission oscillations more apparent.  
The splitting observed in the channels is due to a stronger spin-orbit interaction coupled with valley asymmetry. This interaction produces different effective gaps depending on the spin index, $s_z$, and the valley index, $\tau$. This causes the four channels to propagate with their own wave vector, even though they have the same incoming energy $E^{\tau s_z}$ as confirmed by the wave vectors Eqs. (\ref{kx}-\ref{qx}). 
For a large potential barrier height, only channels with a relatively large wave vector exhibit significant transmission; the others decay rapidly.  The $\bm{K'}\uparrow$ channel exhibits abrupt transmission decay. This behavior is directly related to the smaller wave vector reaching the condition corresponding to an evanescent wave. This makes these particle waves sensitive to slight additions to the barrier potential height. However, $K\downarrow$ corresponds to a higher longitudinal momentum. This gives the particles better penetration through the barrier since they remain propagative for a longer period of time as $V_0$ increases. Consequently, the appearance of the evanescent regime is delayed, and significant transmission is maintained.

\begin{figure}[htp!]
\centering
 \includegraphics[scale=0.46]{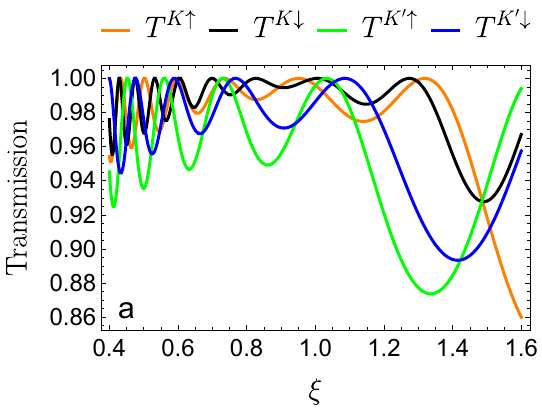}
\includegraphics[scale=0.46]{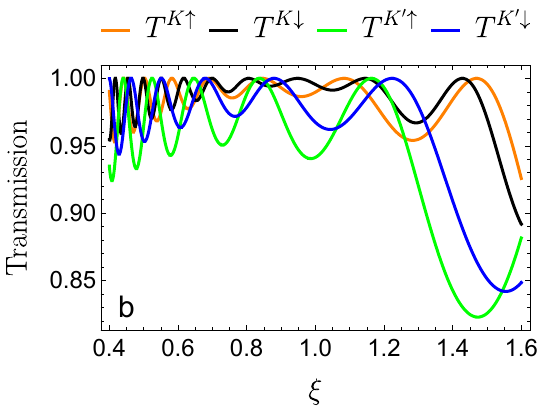}
\includegraphics[scale=0.46]{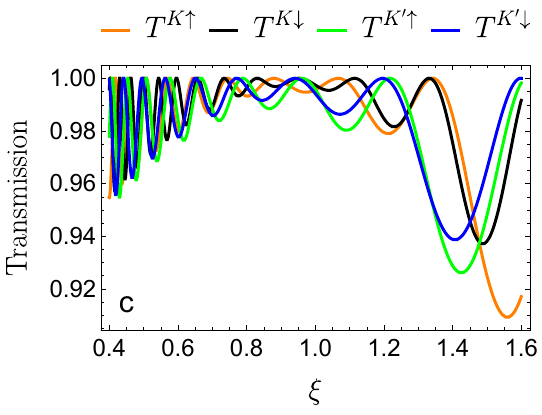}
\caption{\justifying Spin- and valley-dependent transmission  as a function of the velocity ratio $\xi$ 
for $V_0 = 45$ meV, $\phi^{\tau s_z} = 20^\circ$. The incident energy $E^{\tau s_z} = 1.05$ eV for (a): $L = 6$ nm, (b): $L = 7$ nm, and the barrier width $L = 7$ nm for (b): $E^{\tau s_z} = 1.05$ eV, (c): $E^{\tau s_z} = 1.15$ eV.}

\label{T-xi}
\end{figure}


We now examine transmission $T^{\tau s_z}$ versus the velocity ratio $\xi$ in Fig.~\ref{T-xi} for different energy $E^{\tau s_z}$ and barrier width $L$ values. We observe that velocity $v_2$ significantly affects transmission. When  $ \xi \leq 0.8$, the transmission exhibits rapid oscillations with a short period and weak amplitude, particularly for fermions with spin-$\downarrow$ in the $\bm{K}$ valley. As we increase the value of the ratio to around 1.2, the number of these oscillations decreases. The transmission values approach 1 due to strong kinematic matching between regions, resulting in low interface reflectivity.
For sufficiently large values of $\xi$, the wave vector $q_x^{\tau s_z}$  within the barrier decreases, resulting in well-defined resonance conditions with a larger period and amplitude. Meanwhile, transmission values approach the evanescent mode regime. Comparing Fig.~\ref{T-xi}a and Fig.~\ref{T-xi}b clearly shows how increasing $L$ leads to high-amplitude oscillations at large $\xi$, as well as 
{an increased} accumulated phase. However, as the incident energy increases, the total transmission increases, as shown in Fig.~\ref{T-xi}c. This can be explained by the increase in the barrier wave vector, $q_x^{\tau s_z}$. A strong spin-valley dependence is evident in the transmission  $T^{\tau s_z}$. 
The findings show that transmission is controlled. Indeed, the incident energy $E^{\tau s_z}$ determines whether the mode is propagative or quasi-evanescent for the transmitted channels since the barrier allows transmission of high-energy electrons. The barrier width $L$ governs the quantum interference state and the accumulated Fabry-Pérot phase in the potential region. This makes it possible to control the number of oscillations and their periodicity. The Fermi velocity  $v_2$ within the  barrier governs the kinematic matching conditions at the interfaces.

{We close the transmission analysis with the following verification. To confirm the reliability of our transport calculations, we have checked the conservation of the carrier current through the relation $R+T=1$. Since the transmission and reflection probabilities are obtained from the ratios of the transmitted and reflected carrier currents to the incident current, this condition ensures the conservation of the probability flux. For representative values of the incident energy, incidence angle, and barrier parameters considered in Figs.~\ref{T-phi}--\ref{T-xi}, our numerical results confirm that $R+T=1$ within numerical accuracy.}

\begin{figure}[htp!]
\centering
    \includegraphics[scale=0.46]{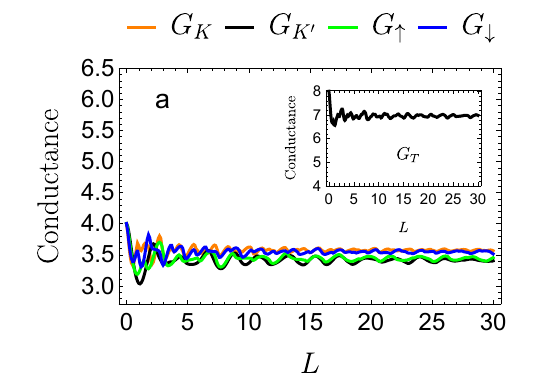}
\includegraphics[scale=0.46]{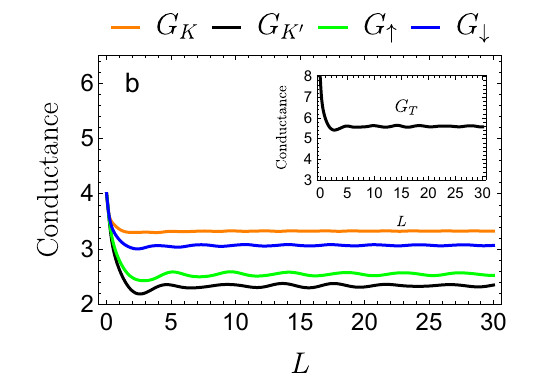}
\includegraphics[scale=0.46]{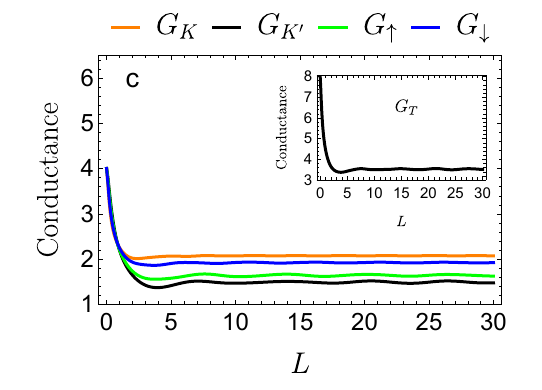}
\caption{\justifying 
	 {The valley-resolved $G_{\tau}$, spin-resolved $G_{s_z}$ conductance} and total conductance $G_T$ as a function of the barrier width  $L$ for $V_0 = 45$ meV,  $E^{\tau s_z} = 1$ eV, and 
three  velocity  ratio values (a): $\xi = 0.5$, (b): $\xi = 1$,  (c): $\xi = 1.5$. }
\label{G-L}
\end{figure}

Figure~\ref{G-L} presents the 
{valley-resolved $G_{\tau}$, spin-resolved $G_{s_z}$ conductance} and the total conductance $G_T$ versus the  barrier width $L$ 
for $V_0 = 45$ meV,  $E^{\tau s_z} = 1$ eV, and 
($\xi = 0.5,1, 1.5$). In Fig.~\ref{G-L}a,  the conductance starts at a value of 
{$G_{\tau} \approx G_{s_z} \approx 4$} and then drops significantly to 
{$G_{\bm{K'}} \approx 3$} for the $\bm{K'}$ valley. This exhibits lower conductance values due to its weak wave vector $q_x^{\bm{K'} s_z}$ and its  energy $E^{\tau s_z}$. As the barrier width $L$ increases, the conductance shows noticeable and irregular oscillations, which are most  pronounced at small values of $L$. 
In contrast, the $\bm{K}$ valley is characterized by a larger wave vector $q_x^{\bm{K}s_z}$ and energy shift as well as high transmission $T^{\bm{K}s_z}$,  which shows a high conductance. At high values of $L$, the conductance of valley $\bm{K'}$ and spin-$\uparrow$ is similar, and valley $\bm{K}$ and spin-$\downarrow$ are also very close to each other. However, this  similarity disappears completely as the velocity $v_2$ inside the barrier increases, as shown in Fig.~\ref{G-L}b.
The conductance becomes smoother with fewer oscillations, particularly for the $\bm{K}$ valley, which displays a consistent value of approximately 3.5. Conversely, the $\bm{K'}$  valley and spin up exhibit a significant decrease. 
The separation of the propagation channels is now more visible. Fig.~\ref{G-L}c shows a drop in conductance 
to a minimum of 1.5 for the $\bm{K'}$ channels and 2.1 for the $\bm{K}$ channels. This can be explained by the small wave vector $q_x^{\tau s_z}$ resulting from the high Fermi velocity $v_2$, a reduction in transmission $T^{\tau s_z}$, and the proximity of the channels to the evanescent regime.
The conductance saturates as the oscillations disappear. The total conductance is larger for $\xi =0.5$, indicating that most modes will propagate inside the barrier with a high wave vector $q_x^{\tau s_z}$. As shown in Fig.~\ref{G-L}b, increasing the  velocity $v_2$ reduces the total conductance depending on the barrier width $L$, reaching saturation at small values of $L$. Similarly, the conductance is suppressed due to the rapid drop observed upon introducing the barrier. A high velocity induces a stronger mismatch, leading to interface reflection and the conversion of propagating modes into evanescent modes.

\begin{figure}[htp!]
	\centering
	 \includegraphics[scale=0.44]{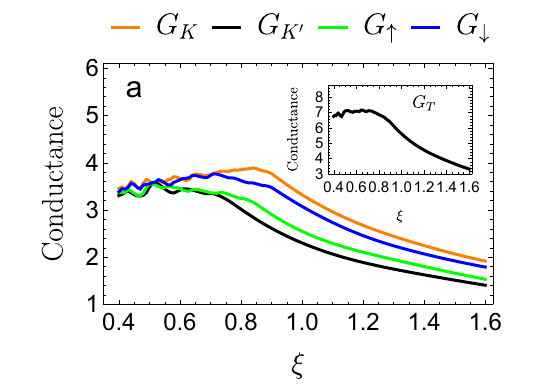}
	\includegraphics[scale=0.44]{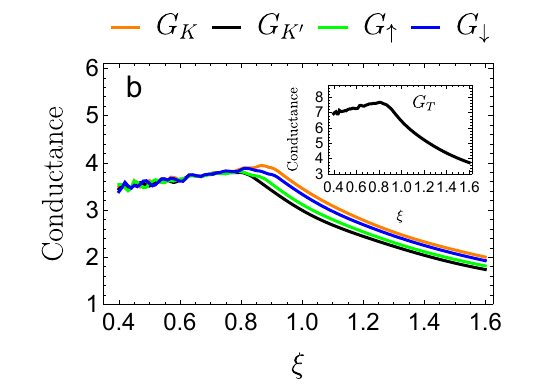}
	\caption{\justifying
		{The valley-resolved $G_{\tau}$, spin-resolved $G_{s_z}$ conductance and total conductance $G_T$} as a function of the velocity  ratio $\xi$ for $V_0 = 45$ meV, $L = 7$ nm, and two energy values
		(a): $E^{\tau s_z} =$ 1 eV, (b): $ E^{\tau s_z} =$ 1.05 eV. }
	\label{G-xi}
\end{figure}

To further emphasize the influence of the  velocity $v_2$ on charge-carrier transport, we analyze the 
 {valley-resolved $G_{\tau}$, spin-resolved $G_{s_z}$ conductance}, versus the velocity ratio $\xi$ in Fig.~\ref{G-xi}, and examine how the incident energy ($E^{\tau s_z}$ = 1, 1.05 eV) impacts the conductance.  
Fig.~\ref{G-xi}a shows the conductance 
for the incident energy $E^{\tau s_z}$ = 1 eV. Examining the conductance at this incoming energy value reveals a clearer dependence on the spin-valley at high Fermi velocity values within the potential barrier. This indicates efficient control of tunneling of fermions in  WSe$_2$.
Fig.~\ref{G-xi}b shows that, for small Fermi velocities $v_2$ (0.4$\leq\xi \leq$0.9), the conductance curves for different spin-$(\uparrow,\downarrow)$ and valley ($\bm{K}$, $\bm{K'}$) are very close, exhibiting small irregular oscillations. However, a significant increase in conductance indicates enhanced carrier transmission. 
	However, when $\xi > 0.9$, the carrier transport regime changes significantly, leading to a progressive decrease in the conductance as $\xi$ increases. The conductance reaches low values
	 {toward the upper end of the considered window, $\xi \to 1.6$}, depending on the spin-$(\uparrow,\downarrow)$ and valley $(\bm{K},\bm{K'})$ indices, because the transport channels become closer to the evanescent regime, as shown in Fig.~\ref{G-L}c. In this region (0.9$\leq\xi\leq$1.6), a clear separation between the different spin-valley channels reveals an effective filtering induced by the barrier.  Furthermore, the total conductance also exhibits a significant increase compared to that in Fig.~\ref{G-xi}a.

\begin{figure}[htp!]
	\centering
	\includegraphics[scale=0.44]{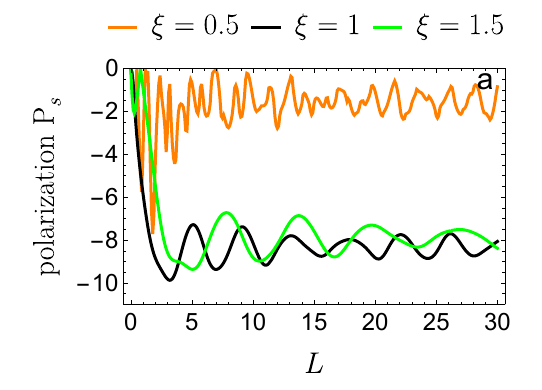}
	\includegraphics[scale=0.44]{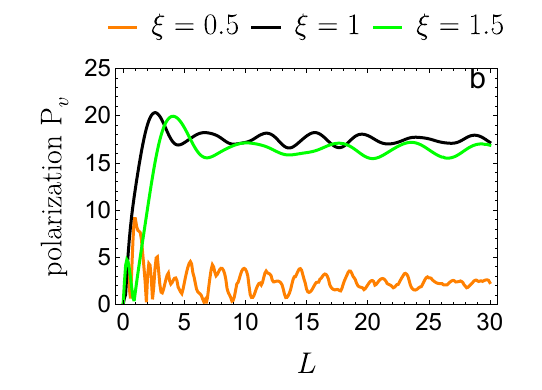} \\
	\includegraphics[scale=0.44]{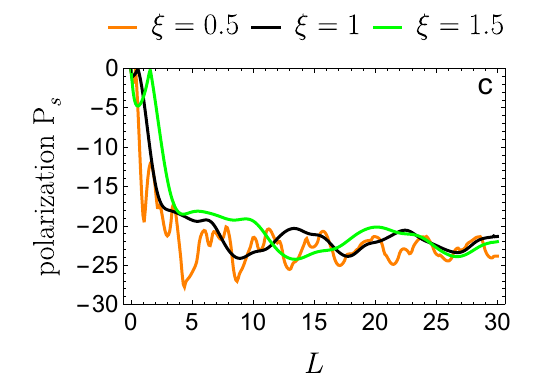}
	\includegraphics[scale=0.44]{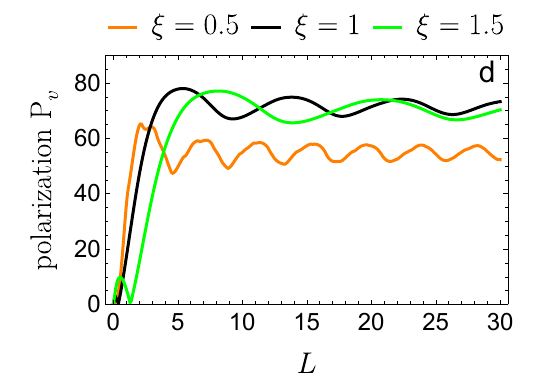}
	\caption{\justifying Spin- and valley-dependent polarization $P_s$ and $P_v$ as a function of the barrier width $L$ for $V_0 = 45$ meV, and two energy values
		(a, b): $E^{\tau s_z} =$ 1 eV,  (c, d): $E^{\tau s_z} =$ 0.95 eV. }
	\label{P-L}
\end{figure}

The analysis in Fig.~\ref{P-L} examines how spin $P_s$ and valley $P_v$ polarizations change with the barrier width $L$ when the barrier height $V_0$ is set to 45 meV and the energy levels are defined as (a,b): $E^{\tau s_z}=$ 1 eV and (c,d): $E^{\tau s_z}=$ 0.95 eV. The spin polarization $P_s$ displays intense unsteady fluctuations, which demonstrate unpredictable patterns according to Fig.~\ref{P-L}a for the case of $\xi = 0.5$. The introduction of $\xi = 1$ and $1.5$ resulted in smoother oscillations, which displayed reduced amplitude while maintaining stable spin polarization.  In Fig.~\ref{P-L}b, the valley polarization $P_v$  displays weak oscillations, which prevent proper valley selection when $\xi = 0.5$. The behavior of $P_v$ at values of $\xi = 1$ and 1.5 exhibits accelerated growth, resulting in a state of equilibrium that facilitates effective valley discrimination. In Fig.~\ref{P-L}c, the spin polarization $P_s$ attains significantly greater negative values compared to Fig.~\ref{P-L}a.  The barrier width increase results in a strong decrease of $P_s$ for all $\xi$ values. The spin-selective suppression showed an increase through this significant reduction of the effect. The valley polarization $P_v$ in Fig.~\ref{P-L}d exhibits substantial growth. The transition from $\xi = 1$ to $\xi = 1.5$  {results} in $P_v$ achieving high stable values throughout various $L$ values while valley polarization maintains a strong presence at $\xi = 0.5$, which demonstrates effective valley filtering. The barrier Fermi velocity increase leads to greater spin-valley selectivity while maintaining stable polarization across multiple barrier width ranges.

\section{Conclusion}
\label{conclusion}

We theoretically investigated spin- and valley-dependent electronic transport in a monolayer transition-metal dichalcogenide, focusing on WSe$_2$, a direct band-gap semiconductor with $\Delta \approx 1.7~\mathrm{eV}$. We considered a simple device geometry consisting of a single rectangular electrostatic barrier. The central mechanism of our study is the control of spin and valley degrees of freedom through Fermi-velocity engineering inside the barrier region, which represents an attractive and experimentally relevant strategy for device-oriented implementations.
Our results demonstrate that the Fermi velocity inside the barrier, $v_2$, acts as a key control parameter. It directly determines the longitudinal wave vector in the confined region and, consequently, the phase accumulated by the wave function across the barrier. As a result, the resonance condition is modified, leading to a shift of the Fabry--Pérot interference pattern generated by multiple reflections within the barrier. The transmission is highly sensitive to changes in $v_2$. Indeed, increasing $v_2$ broadens the spacing between interference fringes, reduces their density, and thus decreases the number of oscillations within a given energy range.

We further computed the spin- and valley-resolved conductance, together with the corresponding spin and valley polarizations, $P_s$ and $P_v$. Our analysis reveals that tuning $v_2$ can strongly enhance both polarizations, even in regimes where the total conductance decreases. This behavior originates from the combined effects of spin--orbit coupling and Zeeman exchange fields, which lift the spin and valley degeneracies and induce channel-dependent shifts of the band edges. Consequently, the longitudinal wave vectors differ for each spin-valley channel. For certain channels, the states become evanescent inside the barrier, while others remain propagating. This selective suppression of specific channels enables efficient spin- and valley-filtered transport.
Overall, the electrostatic barrier behaves as a tunable spin--valley quantum filter. Fermi-velocity engineering provides a direct and efficient means to tailor interference conditions and polarization responses without modifying the geometrical profile of the barrier. This mechanism offers a flexible platform for manipulating internal quantum degrees of freedom in two-dimensional materials. 

From a broader perspective, our findings highlight the potential of velocity engineering in materials with strong spin--orbit coupling as a powerful tool for designing multifunctional nanoelectronic architectures. Such control schemes may be exploited in spintronic and valleytronic devices, including tunable spin--valley filters, polarization-controlled switches, and logic elements based on coupled spin and valley degrees of freedom. These results open new perspectives for the development of next-generation quantum devices based on engineered band-structure and transport properties.


\section*{Acknowledgment}
O. Bouladiane acknowledges the support provided by CNRST in the framework of the program "PhD-Associate Scholarship -- PASS".
H. Bahlouli acknowledges the support provided by the Interdisciplinary Research Center (IRC) for Advanced Materials and King Fahd University of Petroleum \& Minerals (KFUPM).


\begin{thebibliography}{00}
	
	\bibitem{Rita2005} R. C. Iotti, E. Ciancio, and F. Rossi,
	Phys. Rev. B 72, 125347 (2005).
	
	\bibitem{Nicholas1980} J. C. Gill and H. H. Wills,
	Contemp. Phys. 27, 37 (1986).
	
	\bibitem{Martino}A. De Martino, L. Dell'Anna, and R. Egger, Phys. Rev. Lett. 98, 066802 (2007).
	
	\bibitem{Cheol} C.-H. Park, L. Yang, Y.-W. Son, M. L. Cohen, and S. G. Louie,
	Phys. Rev. Lett. 101, 126804 (2008).
	
	\bibitem{2D materials-1}
	N. Benlakhouy, A. El Mouhafid, and A. Jellal,
	Ann. Phys. 468, 169743 (2024).
	
	\bibitem{2D materials-2}S.-H. Ji, J. B. Hannon, R. M. Tromp, V. Perebeinos, J. Tersoff, and F. M. Ross,
	Nat. Mater. 11, 114 (2012).
	
	\bibitem{Novoselov}K. S. Novoselov, A. K. Geim, S. V. Morozov, D. Jiang, Y. Zhang, 
	S. V. Dubonos, I. V. Grigorieva, and A. A. Firsov,
	Science 306, 666 (2004).
	
	\bibitem{Castro2009}A. H. Castro Neto, F. Guinea, N. M. R. Peres, K. S. Novoselov, and A. K. Geim,
	Rev. Mod. Phys. 81, 109 (2009).
	
	\bibitem{approx}Z. Gu, H. A. Fertig, D. P. Arovas, and A. Auerbach, Phys. Rev. Lett. 107, 216601 (2011).
	
	\bibitem{absor}Q. Bao, H. Zhang, B. Wang {\it et al.}, Nat. Photon. 5, 411 (2011).
	
	\bibitem{Novoselov2007}
	K. S. Novoselov, S. V. Morozov, T. M. G. Mohinddin {\it et al.}, Phys. Stat. Sol. (b) 244, 4106 (2007).
	
	\bibitem{Claire Berger} C. Berger, Z. Song, X. Li  {\it et al.}, Science 312, 1191 (2006).
	
	\bibitem{XU} X. Du, I. Skachko, A. Barker, and E. Y. Andrei, Nat. Nanotechnol.  3, 491 (2008).
	
	\bibitem{silicene-1}
	P. Vogt, P. De Padova, C. Quaresima, et al.,
	Phys. Rev. Lett.  108, 155501 (2012).
	
	\bibitem{TMDCs-1}G.-B. Liu, W.-Y. Shan, Y. Yao, W. Yao, and D. Xiao, Phys. Rev. B  88, 085433 (2013).
	
	\bibitem{TMDCs-2}Q. H. Wang, K. Kalantar-Zadeh, A. Kis, J. N. Coleman, and M. S. Strano,
	Nat. Nanotechnol.  7, 699 (2012).
	
	\bibitem{band gap}K. F. Mak, C. Lee, J. Hone, J. Shan, and T. F. Heinz,
	Phys. Rev. Lett.  105, 136805 (2010).
	
	\bibitem{TMDCs-3}M. Chhowalla, Z. Liu, and H. Zhang,
	Chem. Soc. Rev. 44, 2584 (2015).
	
	\bibitem{TMDCs-4}H.-L. Liu, C.-C. Shen, S.-H. Su, C.-L. Hsu, M.-Y. Li, and L.-J. Li,
	Appl. Phys. Lett. 105, 201905 (2014).
	
	\bibitem{TMDCs-5}R. H. Friend and A. D. Yoffe,
	Adv. Phys.  36, 1 (1987).
	
	
	
	\bibitem{transistor-1}B. Radisavljevic, A. Radenovic, J. Brivio, V. Giacometti, and A. Kis,
	Nat. Nanotechnol.  6, 147 (2011).
	
	\bibitem{transistor-2}S. Larentis, B. Fallahazad, and E. Tutuc,
	Appl. Phys. Lett. 101, 223104 (2012).
	
	
	\bibitem{photovoltaics}M. Bernardi, M. Palummo, and J. C. Grossman,
	Nano Lett. 13, 3664 (2013).
	
	\bibitem{WSe2-1}A. Azam, J. Yang, W. Li, J.-K. Huang, and S. Li,
	Prog. Mater. Sci. 132, 101042 (2023).
	
	\bibitem{WSe2-2}A. A. Mitioglu, P. Plochocka, Á. Granados del Aguila {\it et al.},
	Nano Lett. 15, 4387 (2015). 
	
	\bibitem{Zhu2011}Z. Y. Zhu, Y. C. Cheng, and U. Schwingenschl\"ogl,
	Phys. Rev. B 84, 153402 (2011).
	
	\bibitem{Drew2015} D. W. Latzke, W. Zhang, A. Suslu {\it et al.},
	Phys. Rev. B 91, 235202 (2015).
	
	
	\bibitem{valley-1}G. Aivazian, Z. Gong, A. M. Jones {\it et al.},
	Nat. Phys. 11, 148 (2015).
	
	\bibitem{Concha2010}A. Concha and Z. Tešanović,
	Phys. Rev. B 82, 033413 (2010).
	
	\bibitem{Tahir}M. Tahir, P. M. Krstajić, and P. Vasilopoulos,
	Phys. Rev. B 95, 235402 (2017).
	
	\bibitem{PhysRevB.107.245407}D. Xiao, G.-B. Liu, W. Feng, X. Xu, and W. Yao,
	Phys. Rev. Lett. 108, 196802 (2012).
	
	
	
	\bibitem{Srivastava2015}A. Srivastava, M. Sidler, A. V. Allain, D. S. Lembke, A. Kis, and A. Imamoğlu,
	Nat. Phys. 11, 141 (2015).
	
	\bibitem{Zhao2019} C. Zhao, T. Norden, P. Zhang {\it et al.},
	Nat. Nanotechnol. 12, 757 (2017).

	
	

	
	\bibitem{Raoux2010}A. Raoux, M. Polini, R. Asgari, A.R. Hamilton, R. Fazio, A.H. MacDonald, Phys.
	Rev. B 81 (2010) 073407.
	
		\bibitem{Hwang2012}C. Hwang, D. A. Siegel, S.-K. Mo, W. Regan, A. Ismach, Y. Zhang, A. Zettl, and A. Lanzara, Sci. Rep. 2, 590 (2012).
	
	\bibitem{klientun}C. W. J. Beenakker,
	Rev. Mod. Phys. 80, 1337 (2008).
	
	\bibitem{condu0}P. R. Silva, C. Nassif, M. Sampaio, and M. C. Nemes,
	Phys. Lett. A 358, 358 (2006).
	
	\bibitem{butker}M. B\"uttiker, Y. Imry, R. Landauer, and S. Pinhas,
	Phys. Rev. B 31, 6207 (1985).
	
	\bibitem{conduct1}X. Chen and J.-W. Tao,
	Appl. Phys. Lett. 94, 262102 (2009).
	
	
	
	\bibitem{ELAITOUNI2024}R. El Aitouni, M. Mekkaoui, A. Bahaoui, and A. Jellal,
	Appl. Phys. A 131, 48 (2025).
	
	\bibitem{Feng2012}F. Zhai and K. Chang,
	Phys. Rev. B 85, 155415 (2012).
	
	\bibitem{Rekha2025}R. Kumari, G. Dixit, and A. Kundu,
	Phys. Rev. B 111, 155421 (2025).
	
	\bibitem{Moldovan2012}D. Moldovan, M. Ramezani Masir, L. Covaci, and F. M. Peeters,
	Phys. Rev. B 86, 115431 (2012).
	
	\bibitem{X.-J.Hao}X.-J. Hao, R.-Y. Yuan, J.-J. Jin, Y. Guo,
	Front. Phys. 15, 33603 (2020).
	
\end{thebibliography}
\end{document}